\documentclass[a4paper,12pt]{article}
\usepackage{amsthm,amsmath,amssymb,latexsym}

\newtheorem{thm}{Theorem}[section]

\newtheorem{lem}[thm]{Lemma}

\newtheorem{rem}[thm]{Remark}
\numberwithin{equation}{section}

\newcommand{\ds}{\displaystyle}

\newcommand{\wt}{\widetilde}
\newcommand{\wh}{\widehat}
\DeclareMathOperator*{\res}{Res}

\title{The sixth Painlev\'{e} equation arising from $D^{(1)}_4$ hierarchy}
\author{Kenta Fuji and Takao Suzuki\\
{\small Department of Mathematics, Kobe University}\\
{\small Rokko, Kobe 657-8501, Japan}}
\date{}

\begin{document}

\maketitle

\begin{abstract}
The sixth Painlev\'{e} equation arises from a Drinfeld-Sokolov hierarchy of
type $D^{(1)}_4$ by similarity reduction.

\end{abstract}

\section*{Introduction}

The Drinfeld-Sokolov hierarchies are extensions of the KdV (or mKdV) hierarchy
\cite{DS}.
It is known that their similarity reductions imply several Painlev\'{e}
equations \cite{AS,KK1,NY1}.
For the sixth Painlev\'{e} equation ($P_{\mathrm{VI}}$), the relation with the
$A^{(1)}_2$-type hierarchy is investigated \cite{KK2}.
On the other hand, $P_{\mathrm{VI}}$ admits a group of symmetries which is
isomorphic to the affine Weyl group of type $D^{(1)}_4$ \cite{O}.
Also it is known that $P_{\mathrm{VI}}$ is derived from the Lax pair associated
with the algebra $\wh{\mathfrak{so}}(8)$ \cite{NY3}.
However, the relation between $D^{(1)}_4$-type hierarchies and $P_{\mathrm{VI}}$
has not been clarified.
In this paper, we show that the sixth Painlev\'{e} equation is derived from a
Drinfeld-Sokolov hierarchy of type $D^{(1)}_4$ by similarity reduction.

Consider a Fuchsian differential equation on $\mathbb{P}^1(\mathbb{C})$
\begin{equation}\label{Eq:Fuchs}
    \frac{d^2y}{dx^2}+p_1(x)\frac{dy}{dx}+p_2(x)y = 0,
\end{equation}
with the Riemann scheme
\[
	\left\{\begin{array}{cccccc}
		x=t_0& x=t_1& x=t_3& x=t_4& x=\lambda& x=\infty\\[4pt]
		0& 0& 0& 0& 0& \rho\\[4pt]
		\theta_0& \theta_1& \theta_3& \theta_4& 2& \rho+1
	\end{array}\right\},
\]
satisfying the relation
\[
	\theta_0 + \theta_1 + \theta_3 + \theta_4 + 2\rho = 1.
\]
We also let $\mu=\res_{x=\lambda}p_2(x)dx$.
Then the monodromy preserving deformation of the equation \eqref{Eq:Fuchs} is
described as a system of partial differential equations for $\lambda$ and $\mu$.
This system can be regarded as the symmetric representation of
$P_{\mathrm{VI}}$ \cite{Kaw}.
We discuss a derivation of the symmetric representation in the case
\[\begin{array}{lllll}
	\ds t_0=-t& \ds t_1=-\frac{t+1}{t-1}& \ds t_3=\frac{t-1}{t+1}&
	\ds t_4=\frac{1}{t}\\[8pt]
	\theta_0=\alpha_0& \theta_1=\alpha_1-1& \theta_3=\alpha_3-1&
	\theta_4=\alpha_4-1& \rho=\alpha_2.
\end{array}\]
Note that
\[
	\alpha_0 + \alpha_1 + 2\alpha_2 + \alpha_3 + \alpha_4 = 4.
\]
With the notation
\[
	F_0 = \lambda+t,\quad F_1 = \lambda+\frac{t+1}{t-1},\quad F_2 = \mu,\quad
	F_3 = \lambda-\frac{t-1}{t+1},\quad F_4 = \lambda-\frac{1}{t},
\]
the dependence of $\lambda$ and $\mu$ on $t$ is given by
\begin{equation}\begin{split}\label{Sym_P6_1}
	\vartheta(F_j) &= 2F_0F_1F_2F_3F_4 - (\alpha_0-1)F_1F_3F_4\\
	&\quad - (\alpha_1-1)F_0F_3F_4 - (\alpha_3-1)F_0F_1F_4
	- (\alpha_4-1)F_0F_1F_3 + \Theta_j,
\end{split}\end{equation}
for $j=0,1,3,4$ and
\begin{equation}\begin{split}\label{Sym_P6_2}
	\vartheta(F_2) &= -F_2^2(F_0F_1F_3+F_0F_1F_4+F_0F_3F_4+F_1F_3F_4)\\
	&\quad + F_2\{(\alpha_3+\alpha_4-2)F_0F_1+(\alpha_1+\alpha_4-2)F_0F_3
	+(\alpha_1+\alpha_3-2)F_0F_4\\
	&\quad +(\alpha_0+\alpha_4-2)F_1F_3+(\alpha_0+\alpha_3-2)F_1F_4
	+(\alpha_0+\alpha_1-2)F_3F_4\}\\
	&\quad - \alpha_2\{(\alpha_0+\alpha_2-1)F_0+(\alpha_1+\alpha_2-1)F_1
	+(\alpha_3+\alpha_2-1)F_3\\
	&\quad +(\alpha_4+\alpha_2-1)F_4\},
\end{split}\end{equation}
where
\[
	\vartheta = \Theta_0\frac{d}{dt},\quad
	\Theta_i = \prod_{j=0,1,3,4; j\neq i}(F_i-F_j).
\]

Note that the system \eqref{Sym_P6_1}, \eqref{Sym_P6_2} is equivalent to the
Hamiltonian system:
\begin{equation}\label{Ham_P6}
	\frac{d\lambda}{dt} = \frac{\partial H'}{\partial\mu},\quad
	\frac{d\mu}{dt} = -\frac{\partial H'}{\partial\lambda},
\end{equation}
where the Hamiltonian $H'=H'(\lambda,\mu,t)$ is given by
\[\begin{split}
	\Theta_0H' &= F_0F_1F_2^2F_3F_4 - (\alpha_0-1)F_1F_2F_3F_4
	- (\alpha_1-1)F_0F_2F_3F_4\\
	&\quad - (\alpha_3-1)F_0F_1F_2F_4 - (\alpha_4-1)F_0F_1F_2F_3
	+ \alpha_2F_0\{(\alpha_0-1)F_0\\
	&\quad +(\alpha_1+\alpha_2-1)F_1+(\alpha_3+\alpha_2-1)F_3
	+(\alpha_4+\alpha_2-1)F_4\}.
\end{split}\]
We also remark that the system \eqref{Ham_P6} is transformed into the
Hamiltonian system for $P_{\mathrm{VI}}$ as in \cite{IKSY}
\[
	\frac{dq}{ds} = \frac{\partial H}{\partial p},\quad
	\frac{dp}{ds} = -\frac{\partial H}{\partial q},
\]
with the Hamiltonian
\[\begin{split}
	s(s-1)H &= q(q-1)(q-s)p^2 - \frac{1}{4}\{(\alpha_1-4)q(q-1)\\
	&\quad + \alpha_3q(q-s)+\alpha_4(q-1)(q-s)\}p
	+ \frac{1}{16}\alpha_2(\alpha_0+\alpha_2)q,
\end{split}\]
by the canonical transformation $(\lambda,\mu,t,H')\to(q,p,s,H)$ defined as
\[
	q = \frac{(t+\frac{t-1}{t+1})F_4}{(\frac{t-1}{t+1}-\frac{1}{t})F_0},\quad
	p = \frac{(\frac{t-1}{t+1}-\frac{1}{t})F_0(F_0F_2+\alpha_2)}
	{4(t+\frac{t-1}{t+1})(t+\frac{1}{t})},
\]
and
\[
	s = -\frac{(t+\frac{t-1}{t+1})(\frac{t+1}{t-1}+\frac{1}{t})}
	{(t-\frac{t+1}{t-1})(\frac{t-1}{t+1}-\frac{1}{t})}.
\]

This paper is organized as follows.
In Section 1, we recall the definition of the affine Lie algebra
$\mathfrak{g}=\mathfrak{g}(D^{(1)}_4)$.
In Section 2, a Drinfeld-Sokolov hierarchy of type $D^{(1)}_4$ is formulated.
In Sections 3 and 4, we show that its similarity reduction implies the symmetric
representation of $P_{\mathrm{VI}}$.

\section{Affine Lie algebra}

In the notation of \cite{Kac}, the affine Lie algebra
$\mathfrak{g}=\mathfrak{g}(D^{(1)}_4)$ is the Lie algebra generated by the
Chevalley generators $e_i$, $f_i$, $\alpha_i^{\vee}$ $(i=0,\ldots,4)$ and the
scaling element $d$ with the fundamental relations
\[\begin{split}
	&(\mathrm{ad}e_i)^{1-a_{ij}}(e_j)=0,\quad
	(\mathrm{ad}f_i)^{1-a_{ij}}(f_j)=0\quad (i\neq j),\\
	&[\alpha_i^{\vee},\alpha_j^{\vee}]=0,\quad
	[\alpha_i^{\vee},e_j]=a_{ij}e_j,\quad
	[\alpha_i^{\vee},f_j]=-a_{ij}f_j,\quad
	[e_i,f_j]=\delta_{i,j}\alpha_i^{\vee},\\
	&[d,\alpha_i^{\vee}]=0,\quad [d,e_i]=\delta_{i,0}e_0,\quad
	[d,f_i]=-\delta_{i,0}f_0,
\end{split}\]
for $i,j=0,\ldots,4$, where $A=(a_{ij})_{i,j=0}^4$ is the generalized Cartan
matrix of type $D^{(1)}_4$ defined by
\[
	A = \begin{pmatrix}
		2& 0& -1& 0& 0\\
		0& 2& -1& 0& 0\\
		-1& -1& 2& -1& -1\\
		0& 0& -1& 2& 0\\
		0& 0& -1& 0& 2
	\end{pmatrix}.
\]
We denote the Cartan subalgebra of $\mathfrak{g}$ by
\[
	\mathfrak{h} = \bigoplus_{j=0}^{4}\mathbb{C}\alpha_j^{\vee}
	\oplus\mathbb{C}d.
\]
The canonical central element of $\mathfrak{g}$ is given by
\[
	K = \alpha_0^{\vee} + \alpha_1^{\vee} + 2\alpha_2^{\vee} + \alpha_3^{\vee}
	+ \alpha_4^{\vee}.
\]
The normalized invariant form
$(\,|\,):\mathfrak{g}\times\mathfrak{g}\to\mathbb{C}$ is determined by the
conditions
\[\begin{array}{lll}
	(\alpha_i^{\vee}|\alpha_j^{\vee}) = a_{ij},& (e_i|f_j) = \delta_{i,j},&
	(\alpha_i^{\vee}|e_j) = (\alpha_i^{\vee}|f_j) = 0,\\[4pt]
	(d|d) = 0,& (d|\alpha_j^{\vee}) = \delta_{0,j},& (d|e_j) = (d|f_j) = 0,
\end{array}\]
for $i,j=0,\ldots,4$.

We consider the $\mathbb{Z}$-gradation
$\mathfrak{g} = \bigoplus_{k\in\mathbb{Z}}\mathfrak{g}_k(s)$ of type
$s=(1,1,0,1,1)$ by setting
\[\begin{split}
	&\deg\mathfrak{h}=\deg e_2=\deg f_2=0,\\
	&\deg e_i=1,\quad \deg f_i=-1\quad (i=0,1,3,4).
\end{split}\]
If we take an element $d_s\in\mathfrak{h}$ such that
\[
	(d_s|\alpha_2^{\vee}) = 0,\quad (d_s|\alpha_j^{\vee}) = 1\quad (j=0,1,3,4),
\]
this gradation is defined by
\[
	\mathfrak{g}_k(s) = \{x\in\mathfrak{g}\bigm|[d_{s},x]=kx\}\quad
	(k\in\mathbb{Z}).
\]
In the following, we choose
\[
	d_s = 4d + 2\alpha^{\vee}_1 + 3\alpha^{\vee}_2 + 2\alpha^{\vee}_3
	+ 2\alpha^{\vee}_4.
\]
We set
\[
	\mathfrak{g}_{<0} = \bigoplus_{k<0}\mathfrak{g}_{k}(s),\quad
	\mathfrak{g}_{\geq0} = \bigoplus_{k\geq0}\mathfrak{g}_{k}(s).
\]

We choose the graded Heisenberg subalgebra
$\mathfrak{s}=\bigoplus_{k\in\mathbb{Z}}\mathfrak{s}_k(s)$ of $\mathfrak{g}$ of
type $s=(1,1,0,1,1)$ with
\[
	\mathfrak{s}_1(s) = \mathbb{C}\Lambda_{1,1}\oplus\mathbb{C}\Lambda_{1,2},
\]
where
\[\begin{split}
	\Lambda_{1,1} &= -e_0+e_1+e_3-e_{21}+e_{23}+e_{24},\\
	\Lambda_{1,2} &= e_1-e_3+e_4+e_{20}+e_{21}+e_{23}.
\end{split}\]
Here we denote
\[
	e_{2j} = [e_2,e_j],\quad f_{2j} = [f_2,f_j]\quad (j=0,1,3,4).
\]
We remark that
\[
	\mathfrak{s} = \left\{x\in\mathfrak{g}\bigm|[\Lambda_{1,1},x]\in
	\mathbb{C}K\right\}.
\]
and
\[
	\mathfrak{s}_0(s) = \mathbb{C}K,\quad
	\mathfrak{s}_{2k}(s) = 0\quad (k\neq0).
\]
Each $\mathfrak{s}_{2k-1}(s)$ is expressed in the form
\[
	\mathfrak{s}_{2k-1}(s)
	= \mathbb{C}\Lambda_{2k-1,1}\oplus\mathbb{C}\Lambda_{2k-1,2},
\]
with certain elements $\Lambda_{2k-1,i}$ $(i=1,2)$ satisfying
\[
	[\Lambda_{2k-1,i},\Lambda_{2l-1,j}]
	= (2k-1)\delta_{i,j}\delta_{k+l,1}K\quad (i,j=1,2; k,l\in\mathbb{Z}).
\]
For $k=0$, we have
\[\begin{split}
	\Lambda_{-1,1} &= \frac{1}{2}(-2f_0+f_1+f_3+f_{21}-f_{23}-2f_{24}),\\
	\Lambda_{-1,2} &= \frac{1}{2}(f_1-f_3+2f_4-2f_{20}-f_{21}-f_{23}).
\end{split}\]

\begin{rem}
In the nortation of {\rm\cite{C}}, the Heisenberg subalgebra $\mathfrak{s}$
corresponds to the conjugacy class $D_4(a_1)$ of the Weyl group $W(D_4)${\rm;}
see {\rm\cite{DF}}.
\end{rem}

\section{Drinfeld-Sokolov hierarchy}

In the following, we use the notation of infinite dimensional groups
\[
	G_{<0} = \exp(\wh{\mathfrak{g}}_{<0}),\quad
	G_{\geq0} = \exp(\wh{\mathfrak{g}}_{\geq0}),
\]
where $\wh{\mathfrak{g}}_{<0}$ and $\wh{\mathfrak{g}}_{\geq0}$ are completions
of $\mathfrak{g}_{<0}$ and $\mathfrak{g}_{\geq0}$ respectively.

Introducing the time variables $t_{k,i}$ $(i=1,2; k=1,3,5,\ldots)$, we consider
the {\it Sato equation} for a $G_{<0}$-valued function
$W=W(t_{1,1},t_{1,2},\ldots)$
\begin{equation}\label{Eq:Sato}
	\partial_{k,i}(W) = B_{k,i}W - W\Lambda_{k,j}\quad (i=1,2; k=1,3,5,\ldots),
\end{equation}
where $\partial_{k,i}=\partial/\partial t_{k,i}$ and $B_{k,i}$ stands for the
$\mathfrak{g}_{\geq0}$-component of
$W\Lambda_{k,i}W^{-1}\in\wh{\mathfrak{g}}_{<0}\oplus\mathfrak{g}_{\geq0}$.
We understand the Sato equation \eqref{Eq:Sato} as a conventional form of the
differential equation
\begin{equation}\label{Eq:Sato2}
	\partial_{k,i} - B_{k,i} = W(\partial_{k,i}-\Lambda_{k,i})W^{-1}\quad
	(i=1,2; k=1,3,5,\ldots),
\end{equation}
defined through the adjoint action of $G_{<0}$ on
$\wh{\mathfrak{g}}_{<0}\oplus\mathfrak{g}_{\geq0}$.
The Zakharov-Shabat equation
\begin{equation}\label{ZS_DS}
	[\partial_{k,i}-B_{k,i},\partial_{l,j}-B_{l,j}] = 0\quad
	(i,j=1,2; k,l=1,3,5,\ldots),
\end{equation}
follows from the Sato equation \eqref{Eq:Sato2}.

The $\mathfrak{g}_{\geq0}$-valued functions $B_{1,i}$ $(i=1,2)$ are expressed in
the form
\begin{equation}\label{DS_depvar_def}
	B_{1,i} = \Lambda_{1,i} + U_i,\quad
	U_i = \sum_{j=0}^{4}u_{j,i}\alpha_j^{\vee} + x_ie_2 + y_if_2.
\end{equation}
The Zakharov-Shabat equation \eqref{ZS_DS} for $k=1$ is equivalent to
\begin{equation}\label{ZS_U}
	\partial_{1,i}(U_j) - \partial_{1,j}(U_i) + [U_j,U_i] = 0,\quad
	[\Lambda_{1,i},U_j] - [\Lambda_{1,j},U_i] = 0,
\end{equation}
for $i,j=1,2$.
Then we have
\begin{lem}\label{Lem:U_Rel}
Under the Sato equation \eqref{Eq:Sato2}, the following equations are
satisfied{\rm:}
\begin{equation}\label{ZS_U_grad}
	(d_s|\partial_{1,i}(U_j)) + \frac{1}{2}(U_i|U_j) = 0\quad (i,j=1,2).
\end{equation}
\end{lem}

\begin{proof}
The system \eqref{Eq:Sato2} for $k=1$ is equivalent to
\begin{equation}\label{Eq:S-W}
	\partial_{1,i} - \Lambda_{1,i} - U_i
	= W(\partial_{1,i}-\Lambda_{1,i})W^{-1}\quad (i=1,2).
\end{equation}
Set
\[
	W = \exp(w),\quad w = \sum_{k=1}^{\infty}w_{-k},\quad
	w_{-k}\in\mathfrak{g}_{-k}(s).
\]
Then the system \eqref{Eq:S-W} implies
\begin{equation}\label{Eq:S-W_ad}
	U_i = \sum_{k=1}^{\infty}\frac{1}{k!}\mathrm{ad}(w)^{k-1}\partial_{1,i}(w)
	+ \sum_{k=1}^{\infty}\frac{1}{k!}\mathrm{ad}(w)^k(\Lambda_{1,i})\quad
	(i=1,2).
\end{equation}
Comparing the component of degree $-k$ in \eqref{Eq:S-W_ad}, we obtain
\[
	U_i = \mathrm{ad}(w_{-1})(\Lambda_{1,i})\quad (i=1,2),
\]
for $k=0$;
\begin{equation}\begin{split}\label{Eq:S-W_ad_d1}
	\mathrm{ad}(w_{-2})(\Lambda_{1,i})
	+ \frac{1}{2}\mathrm{ad}(w_{-1})^2(\Lambda_{1,i})
	+ \partial_{1,i}(w_{-1}) = 0\quad (i=1,2),
\end{split}\end{equation}
for $k=1$;
\[\begin{split}
	&\sum_{i_1+\ldots+i_l=k+1}\frac{1}{l!}\mathrm{ad}(w_{-i_1})\ldots
	\mathrm{ad}(w_{-i_l})(\Lambda_{1,i})\\
	&\qquad + \sum_{i_1+\ldots+i_l=k}\frac{1}{l!}\mathrm{ad}(w_{-i_1})\ldots
	\mathrm{ad}(w_{-i_{l-1}})\partial_{1,i}(w_{-i_l}) = 0\quad (i=1,2),
\end{split}\]
for $k\geq2$.
On the other hand, we have
\[
	(\Lambda_{1,i}|\mathrm{ad}(\Lambda_{1,j})(x)) = 0\quad
	(i,j=1,2; x\in\mathfrak{g}_{-2}(s)),
\]
and
\[
	(\Lambda_{1,i}|x) = (d_s|\mathrm{ad}(\Lambda_{1,i})(x))\quad
	(i=1,2; x\in\mathfrak{g}_{-1}(s)).
\]
Hence it follows that
\[\begin{split}
	(\Lambda_{1,j}|\textrm{LHS of \eqref{Eq:S-W_ad_d1}})
	&= \frac{1}{2}(\Lambda_{1,j}|\mathrm{ad}(w_{-1})^2(\Lambda_{1,i}))
	+ (\Lambda_{1,j}|\partial_{1,i}(w_{-1}))\\
	&= -\frac{1}{2}(U_i|U_j) - (d_s|\partial_{1,i}(U_j)).
\end{split}\]
\end{proof}

\begin{rem}
Let $X(0)\in G_{<0}G_{\geq0}$ and define
\[
	X = X(t_{1,1},t_{1,2},\ldots) = \exp(\xi)X(0),\quad
	\xi = \sum_{i=1,2}\sum_{k=1,3,\ldots}t_{k,i}\Lambda_{k,i}.
\]
Then a solution $W\in G_{<0}$ of the system \eqref{Eq:Sato} is given formally
via the decomposition
\[
	X = W^{-1}Z,\quad Z\in G_{\geq0}.
\]
\end{rem}

\section{Similarity reduction}

Under the Sato equation \eqref{Eq:Sato2}, we consider the operator
\[
	\mathcal{M} = W\exp(\xi)d_s\exp(-\xi)W^{-1},\quad
	\xi = \sum_{i=1,2}\sum_{k=1,3,\ldots}t_{k,i}\Lambda_{k,i}.
\]
Then the operator $\mathcal{M}$ satisfies
\[
	\partial_{k,i}(\mathcal{M}) = [B_{k,i},\mathcal{M}]\quad
	(i=1,2; k=1,3,5,\ldots).
\]
Note that
\[
	\mathcal{M} = d_s - \sum_{i=1,2}\sum_{k=1,3,\ldots}
	kt_{k,i}W\Lambda_{k,i}W^{-1} - d_s(W)W^{-1}.
\]

Assuming that $t_{k,1}=t_{k,2}=0$ for $k\geq3$, we require that the similarity
condition $\mathcal{M}\in\mathfrak{g}_{\geq0}$ is satisfied.
Then we have
\[
	\partial_{1,i}(\mathcal{M}) = [B_{1,i},\mathcal{M}]\quad (i=1,2).
\]
where $\mathcal{M}=d_s-t_{1,1}B_{1,1}-t_{1,2}B_{1,2}$, or equivalently
\begin{equation}\label{Sim_Red}
	[d_s-M,\partial_{1,i}-B_{1,i}] = 0\quad (i=1,2),
\end{equation}
where $M=t_{1,1}B_{1,1}+t_{1,2}B_{1,2}$.
Under the Zakharov-Shabat equation
\[
	[\partial_{1,1}-B_{1,1},\partial_{1,2}-B_{1,2}] = 0,
\]
the system \eqref{Sim_Red} is equivalent to
\[
	\sum_{j=1,2}t_{1,j}\partial_{1,j}(B_{1,i}) = [d_s,B_{1,i}]-B_{1,i}\quad
	(i=1,2).
\]
In terms of the operators $U_i$, this similarity condition can be
expressed as
\begin{equation}\label{Sim_Red_U}
	\sum_{j=1,2}t_{1,j}\partial_{1,j}(U_i) + U_i = 0\quad (i=1,2).
\end{equation}
We regard the systems \eqref{ZS_U}, \eqref{ZS_U_grad} and \eqref{Sim_Red_U} as
a similarity reduction of the Drinfeld-Sokolov hierarchy of type $D^{(1)}_4$.

In the notation \eqref{DS_depvar_def}, these systems are expressed in terms of
the variables $u_{j,i}$, $x_i$, $y_i$ as follows:
\[\begin{split}
	&\partial_{1,1}(x_2) - \partial_{1,2}(x_1)\\
	&\quad - (u_{1,1}-u_{3,1}-u_{0,2}+u_{4,2})x_1
	+ (u_{0,1}-u_{4,1}+u_{1,2}-u_{3,2})x_2 = 0,\\
	&\partial_{1,1}(y_2) - \partial_{1,2}(y_1)\\
	&\quad + (u_{1,1}-u_{3,1}-u_{0,2}+u_{4,2})y_1
	- (u_{0,1}-u_{4,1}+u_{1,2}-u_{3,2})y_2 = 0,\\
	&\partial_{1,1}(u_{2,2}) - \partial_{1,2}(u_{2,1}) - x_1y_2 + x_2y_1 = 0,\\
	&\partial_{1,1}(u_{j,2}) - \partial_{1,2}(u_{j,1}) = 0\quad (j=0,1,3,4),
\end{split}\]
and
\begin{equation}\begin{split}\label{DS_depvar_1}
	&u_{1,1} - 2u_{2,1} + u_{3,1} + 2u_{4,1} - u_{1,2} + u_{3,2} = 0,\\
	&u_{1,1} - u_{3,1} - 2u_{0,2} - u_{1,2} + 2u_{2,2} - u_{3,2} = 0,\\
	&u_{1,1} - u_{3,1} + u_{1,2} + u_{3,2} - 2u_{4,2} + 2x_1 = 0,\\
	&2u_{0,1} - u_{1,1} - u_{3,1} - u_{1,2} + u_{3,2} + 2x_2 = 0,\\
	&u_{1,1} - u_{3,1} + 2u_{0,2} - u_{1,2} - u_{3,2} + 2y_1 = 0,\\
	&u_{1,1} + u_{3,1} - 2u_{4,1} - u_{1,2} + u_{3,2} + 2y_2 = 0,\\
\end{split}\end{equation}
for the system \eqref{ZS_U};
\[\begin{split}
	&\sum_{l=0,1,3,4}4\partial_{1,i}(u_{l,j})\\
	&\quad + \sum_{l=0,1,3,4}(2u_{l,i}-u_{2,i})(2u_{l,j}-u_{2,j})
	+ 2(x_iy_j+y_ix_j) = 0\quad (i,j=1,2),
\end{split}\]
for the system \eqref{ZS_U_grad};
\[\begin{split}
	&t_{1,1}\partial_{1,1}(x_i) + t_{1,2}\partial_{1,2}(x_i) + x_i = 0,\quad
	t_{1,1}\partial_{1,1}(y_i) + t_{1,2}\partial_{1,2}(y_i) + y_i = 0,\\
	&t_{1,1}\partial_{1,1}(u_{j,i}) + t_{1,2}\partial_{1,2}(u_{j,i})
	+ u_{j,i} = 0,\quad (i=1,2; j=0,\ldots,4),
\end{split}\]
for the system \eqref{Sim_Red_U}.
In the next section, we show that they imply the sixth Painlev\'{e} equation.

Under the similarity condition \eqref{Sim_Red_U}, the system \eqref{ZS_U_grad}
implies
\[
	2(d_s|U_i) - t_{1,1}(U_i|U_1) - t_{1,2}(U_i|U_2) = 0\quad (i=1,2).
\]
It is expressed in terms of the variables $u_{j,i}$, $x_i$, $y_i$ as follows:
\begin{equation}\begin{split}\label{DS_depvar_2}
	&\sum_{l=0,1,3,4}4u_{l,i}
	- \sum_{l=0,1,3,4}t_{1,1}(2u_{l,i}-u_{2,i})(2u_{l,1}-u_{2,1})
	- 2t_{1,1}(x_iy_1+y_ix_1)\\
	&\quad - \sum_{l=0,1,3,4}t_{1,2}(2u_{l,i}-u_{2,i})(2u_{l,2}-u_{2,2})
	- 2t_{1,2}(x_iy_2+y_ix_2) = 0\quad (i=1,2).
\end{split}\end{equation}

\begin{rem}
The systems \eqref{ZS_U} and \eqref{Sim_Red_U} can be regarded as the
compatibility condition of the Lax form
\begin{equation}\label{Lax_Sim_Red}
	d_s(\Psi) = M\Psi,\quad \partial_{1,i}(\Psi) = B_{1,i}\Psi\quad (i=1,2),
\end{equation}
where $\Psi = W\exp(\xi)$.
\end{rem}

\section{The sixth Painlev\'{e} equation}

In the previous section, we have derived the system of the equations
\begin{equation}\begin{split}\label{DS_U}
	&\partial_{1,i}(U_j) - \partial_{1,j}(U_i) + [U_j,U_i] = 0,\quad
	[\Lambda_{1,i},U_j] - [\Lambda_{1,j},U_i] = 0,\\
	&(d_s|\partial_{1,i}(U_j)) - \frac{1}{2}(U_i|U_j) = 0,\quad
	\sum_{l=1,2}t_{1,l}\partial_{1,l}(U_i) + U_i = 0\quad (i,j=1,2),
\end{split}\end{equation}
for the $\mathfrak{g}_0$-valued functions $U_i=U_i(t_{1,1},t_{1,2})$ $(i=1,2)$,
as a similarity reduction of the $D^{(1)}_4$ hierarchy of type $s=(1,1,0,1,1)$.
In terms of the operators $B_{1,i}=\Lambda_{1,i}+U_i$ and
$M=t_{1,1}B_{1,1}+t_{1,2}B_{1,2}$, the system \eqref{DS_U} is expressed as
\[
	[\partial_{1,1}-B_{1,1},\partial_{1,2}-B_{1,2}] = 0,\quad
	[d_s-M,\partial_{1,i}-B_{1,i}] = 0\quad (i=1,2),
\]
with the equations for normalization \eqref{ZS_U_grad}.
In this section, we show that the sixth Painlev\'{e} equation is derived from
them.

The operator $M$ is expressed in the form
\[\begin{split}
	M = \sum_{i=1,2}t_{1,i}\Lambda_{1,i}
	+ \sum_{j=0,1,3,4}\kappa_j\alpha_j^{\vee} + \eta\alpha_2^{\vee}
	+ \varphi e_2 + \psi f_2,
\end{split}\]
so that
\begin{equation}\begin{split}\label{DS_depvar_3}
	&\kappa_j = t_{1,1}u_{j,1} + t_{1,2}u_{j,2}\quad (j=0,1,3,4),\quad
	\eta = t_{1,1}u_{2,1} + t_{1,2}u_{2,2},\\
	&\varphi = t_{1,1}x_1 + t_{1,2}x_2,\quad \psi = t_{1,1}y_1 + t_{1,2}y_2.
\end{split}\end{equation}
The system \eqref{Sim_Red} implies that the variables $\kappa_j$ $(j=0,1,3,4)$
are independent of $t_{1,i}$ $(i=1,2)$.
Then the following lemma is obtained from \eqref{DS_depvar_1},
\eqref{DS_depvar_2} and \eqref{DS_depvar_3}.
\begin{lem}\label{Lem:DS_depvar}
The variables $u_{j,i}$, $x_i$, $y_i$ $(i=1,2; j=0,\ldots,4)$ are determined
uniquely as polynomials in $\eta$, $\varphi$ and $\psi$ with coefficients in
$\mathbb{C}(t_{1,i})[\kappa_j]$.
Furthermore, the following relation is satisfied{\rm:}
\[
	\eta^2 - (\kappa_0+\kappa_1+\kappa_3+\kappa_4)(\eta+1)
	+ \kappa_0^2 + \kappa_1^2 + \kappa_3^2 + \kappa_4^2 + \varphi\psi = 0.
\]
\end{lem}
Thanks to this lemma, the system \eqref{DS_U} can be rewritten into a system of
first order differential equations for $\eta$ and $\varphi$; we do not give
the explicit formulas here.

We denote by $\mathfrak{n}_{+}$ the subalgebra of $\mathfrak{g}$ generated by
$e_j$ $(j=0,\ldots,4)$, and by $\mathfrak{b}_{+}$ the borel subalgebra of $\mathfrak{g}$ defined by $\mathfrak{b}_{+}=\mathfrak{h}\oplus\mathfrak{n}_{+}$.
We look for a dependent variable $\lambda$ such that
\[\begin{split}
	\wt{M} &= \exp(-\lambda f_2)M\exp(\lambda f_2)
	- \exp(-\lambda f_2)d_s(\exp(\lambda f_2))\in\mathfrak{b}_{+},\\
	\wt{B}_{1,i} &= \exp(-\lambda f_2)B_{1,i}\exp(\lambda f_2)
	- \exp(-\lambda f_2)\partial_{1,i}(\exp(\lambda f_2))
	\in\mathfrak{b}_{+}\quad (i=1,2),
\end{split}\]
namely
\begin{equation}\begin{split}\label{Eq:Lambda_def}
	&\varphi\lambda^2 + (2\eta-\kappa_0-\kappa_1-\kappa_3-\kappa_4)\lambda
	- \psi = 0,\\
	&\partial_{1,i}(\lambda) + x_i\lambda^2
	- (u_{0,i}+u_{1,i}-2u_{2,i}+u_{3,i}+u_{4,i})\lambda - y_i = 0\quad (i=1,2).
\end{split}\end{equation}
Note that the definition of $\wt{M}$ and $\wt{B}_{1,i}$ arises from the gauge
transformation $\Psi\to\Phi$ defined by $\Phi=\exp(-\lambda f_2)\Psi$ on the Lax
form
\eqref{Lax_Sim_Red}.
By Lemma \ref{Lem:DS_depvar} together with the system \eqref{DS_U}, we can show
that
\[
	\lambda = -\frac{1}{8\varphi}
	(8\eta-\alpha_0^2-\alpha_1^2-\alpha_3^2-\alpha_4^2+4),
\]
satisfies the equation \eqref{Eq:Lambda_def}, where $\alpha_j$ $(j=0,1,3,4)$
are constants defined by
\[
	\kappa_j
	= -\frac{1}{16}(8\alpha_j-\alpha_0^2-\alpha_1^2-\alpha_3^2-\alpha_4^2-4).
\]
We also let $\mu$ by a dependent variable defined by $\mu=\varphi$ so that
\[
	\eta = -\lambda\mu
	+ \frac{1}{8}(\alpha_0^2+\alpha_1^2+\alpha_3^2+\alpha_4^2-4),\quad
	\varphi = \mu.
\]
Then the system \eqref{DS_U} can be regarded as a system of differential
equations for variables $\lambda$ and $\mu$ with parameters $\alpha_j$
$(j=0,1,3,4)$.

We now regard the system \eqref{DS_U} as a system of ordinary differential
equations with respect to the independent variable $t=t_{1,1}$ by setting
$t_{1,2}=1$.
Then the operator $\wt{M}$ is written in the form
\[\begin{split}
	\wt{M} &= \frac{1}{16}(\alpha_0^2+\alpha_1^2+\alpha_3^2+\alpha_4^2-4)K
	- \sum_{j=0,1,3,4}\frac{1}{2}(\alpha_j-1)\alpha_j^{\vee}\\
	&\quad + F_2e_2 - F_0e_0 + (t-1)F_1e_1 - (t+1)F_3e_3 - tF_4e_4\\
	&\quad + e_{20} - (t-1)e_{21} + (t+1)e_{23} + te_{24},
\end{split}\]
where
\[
	F_0 = \lambda+t,\quad F_1 = \lambda+\frac{t+1}{t-1},\quad F_2 = \mu,\quad
	F_3 = \lambda-\frac{t-1}{t+1},\quad F_4 = \lambda-\frac{1}{t}.
\]
The operator $\wt{B}=\wt{B}_{1,1}$  is written in the form
\[\begin{split}
	\wt{B} &= \wt{u}_2K + \sum_{j=0,1,3,4}\wt{u}_j\alpha_j^{\vee} + \wt{x}e_2\\
	&\quad - e_0 + (\lambda+1)e_1 - (\lambda-1)e_3 - \lambda e_4 - e_{21}
	+ e_{23} + e_{24},
\end{split}\]
where $\wt{u}_2$ is a polynomial in $\lambda$, $\mu$ and the other coefficients
are given by
\[\begin{split}
	\Theta_0\wt{u}_j &= F_0F_1F_2F_3F_4F_j^{-1} - \sum_{i=0,1,3,4; i\neq j}
	\frac{1}{2}(\alpha_i+\alpha_j-2)F_0F_1F_3F_4F_i^{-1}F_j^{-1}\\
	&\quad - \frac{1}{2}(\alpha_j-1)F_0(F_0-F_1-F_3-F_4)\quad
	(j=0,1,3,4),\\[4pt]
	\Theta_0\wt{x} &= F_0F_2(F_0-F_1-F_3-F_4) + (\alpha_0+\alpha_2-1)F_0\\
	&\quad + (\alpha_1+\alpha_2-1)F_1 + (\alpha_3+\alpha_2-1)F_3
	+ (\alpha_4+\alpha_2-1)F_4,
\end{split}\]
with
\[
	\Theta_0 = (F_0-F_1)(F_0-F_3)(F_0-F_4),\quad
	\alpha_2 = -\frac{1}{2}(\alpha_0+\alpha_1+\alpha_3+\alpha_4-1).
\]
Since $\wt{M}$ and $\wt{B}$ is obtained from $M$ and $B_{1,1}$ by the gauge
transformation, they satisfy
\[
	\left[d_s-\wt{M},\frac{d}{dt}-\wt{B}\right] = 0.
\]
By rewriting this compatibility condition into differential equations for $F_j$
$(j=0,\ldots,4)$, we obtain the same system as \eqref{Sym_P6_1},
\eqref{Sym_P6_2}.
\begin{thm}
Under the specialization $t_{1,1}=t$ and $t_{1,2}=1$, the system \eqref{DS_U}
is equivalent to the sixth Painlev\'{e} equation \eqref{Sym_P6_1},
\eqref{Sym_P6_2}.
\end{thm}

\begin{rem}
The system \eqref{Sym_P6_1}, \eqref{Sym_P6_2} can be regarded as the
compatibility condition of the Lax pair
\begin{equation}\label{Lax_P6}
	d_s(\Phi) = \wt{M}\Phi,\quad \frac{d\Phi}{dt} = \wt{B}\Phi,
\end{equation}
where $\Phi=\exp(-\lambda f_2)W\exp(\xi)$.
Let
\[\begin{split}
	\Omega &= \exp(\omega_1\alpha_1^{\vee}+\omega_2\alpha_2^{\vee}
	+\omega_3\alpha_3^{\vee}+\omega_4\alpha_4^{\vee})\exp(F_0^{-1}e_2)\Phi,
\end{split}\]
where
\[\begin{array}{ll}
	\ds\omega_1 = \frac{1}{2}\log(t^2+2t-1)(t^2+1),&
	\ds\omega_2 = \log F_0,\\[8pt]
	\ds\omega_3 = \frac{1}{2}\log(1+2t-t^2)(t^2+1),&
	\ds\omega_4 = \frac{1}{2}\log(1+2t-t^2)(t^2+2t-1).
\end{array}\]
Then the system \eqref{Lax_P6} is transformed into the Lax pair of the type of
{\rm\cite{NY3}} by the gauge transformation $\Phi\to\Omega$.
\end{rem}

Finally, we define the group of symmetries for $P_{\mathrm{VI}}$ following
\cite{NY2}.
Consider the transformations
\[
	r_i(X) = X\exp(-e_i)\exp(f_i)\exp(-e_i)\quad (i=0,\ldots,4),
\]
where
\[
	X = \exp(\xi)X(0) = W^{-1}Z,\quad
	\xi = \sum_{i=1,2}\sum_{k=1,3,\ldots}t_{k,i}\Lambda_{k,i}.
\]
Under the similarity condition
$\mathcal{M}\in\mathfrak{g}_{\geq0}$, their action on $W$ is given by
\[\begin{split}
	r_i(W) &= \exp(\lambda f_2)
	\exp\left(\frac{(\alpha_i^{\vee}|d_s-\wt{M})}{(f_i|d_s-\wt{M})}f_i\right)
	\exp(-\lambda f_2)W\quad (i=0,1,3,4),\\
	r_2(W) &= W.
\end{split}\]
We also define
\[
	r_i(\alpha_j) = \alpha_j - \alpha_ia_{ij}\quad (i,j=0,\ldots,4).
\]
Then the action of them on the variables $\lambda$, $\mu$ is described as
\[
	r_i(F_j) = F_j - \frac{\alpha_i}{F_i}u_{ij}\quad (i,j=0,\ldots,4),
\]
where $U=(u_{ij})_{i,j=0}^{4}$ is the orientation matrix of the Dynkin diagram
defined by
\[
	U = \begin{pmatrix}
		0& 0& 1& 0& 0\\
		0& 0& 1& 0& 0\\
		-1& -1& 0& -1& -1\\
		0& 0& 1& 0& 0\\
		0& 0& 1& 0& 0
	\end{pmatrix}.
\]
Note that the transformations $r_i$ $(i=0,\ldots,4)$ satisfy the fundamental
relations for the generators of the affine Weyl group $W(D^{(1)}_4)$.

\section*{Acknowledgement}
The authers are grateful to Professors Masatoshi Noumi, Yasuhiko Yamada, Saburo
Kakei and Tetsuya Kikuchi for valuable discussions and advices.


\end{document}